\pdfoutput=1

\documentclass[8pt, twoside]{article} 
\usepackage{times}
\usepackage{graphicx}
\usepackage[rflt]{floatflt}
\usepackage[font={small}]{caption}
\usepackage{caption3}
\usepackage{subcaption}

\usepackage{color}
\definecolor{hellgrau}{rgb}{0.902,0.902,0.902}
\definecolor{hellorange}{rgb}{1.0, 0.937, 0.8}
\definecolor{navyblue}{rgb}{0.098, 0.110, 0.576}

\renewcommand{\baselinestretch}{1.0}

\usepackage{fancyhdr}
\begin{document}

\setlength{\parindent}{0cm}
\normalsize

\renewcommand{\baselinestretch}{1.0}\normalsize

\twocolumn[
{\large \colorbox{navyblue}{\color{white}  \textbf{BIOPHYSICS}}\\*[0.4cm]}
{\Huge \textbf{The Physics of Nerves}\\*[0.1cm]}
{\large \color{navyblue} Physical concepts help describing the propagation of nerve pulses$^\ast$\\*[0.4cm]}
{\large \textbf{Thomas Heimburg} - Niels Bohr Institute, University of Copenhagen, Denmark\\*[0.5cm]}
]


\let\thefootnote\relax\footnotetext{$^{\ast}$The German language original was published in the \textit{Physik Journal} of the German Physical Society (DPG) in 2009. E-mail: theimbu@nbi.dk}  
\textbf{The accepted model for nerve pulse propagation in biological membranes seems insufficient. It is restricted to dissipative electrical phenomena and considers nerve pulses exclusively as a microscopic phenomenon.  A simple thermodynamic model that is based on the macroscopic properties of membranes allows explaining more features of nerve pulse propagation including the phenomenon of anesthesia that has so far remained unexplained.}\\

In his book entitled \textit{"What is life?"} from 1944 Erwin Schr\"odinger summarized his view of the role that physics plays in biology. He outlined his opinion that the laws of physics, and in particular those of thermodynamics, are sufficient to explain life. His book had a big influence on his contemporary colleagues. James Watson and Francis Crick, the finders of the DNA double helix structure, explicitly referred to it. Schr\"odinger based his considerations on the work of Max Delbr\"uck who began his career as a quantum mechanist as an assistant of Lise Meitner in Berlin. Encouraged by Niels Bohr he turned to biological problems in the 1930s. Thanks to his research on bacteriophages he became one of the fathers of modern molecular biology. James Watson was one of his students. In spite of the prominent role that physicists played during the development of modern molecular biology this discipline is strangely alien to them. Familiar functions as temperature, pressure, heat and entropy only seem to play a secondary role, or no role at all. Instead, one finds complex reaction networks made up of thousands of molecules which, due to their sheer quantity, partially have no names but rather abbreviations and numbering. This approach is hardly accessible to many physicists, and this may partially be responsible for the fact that biophysics as a scientific discipline within physics was only very slowly accepted by universities.\\
\begin{figure}[ht!]
    \begin{center}
	\includegraphics[width=8.5cm]{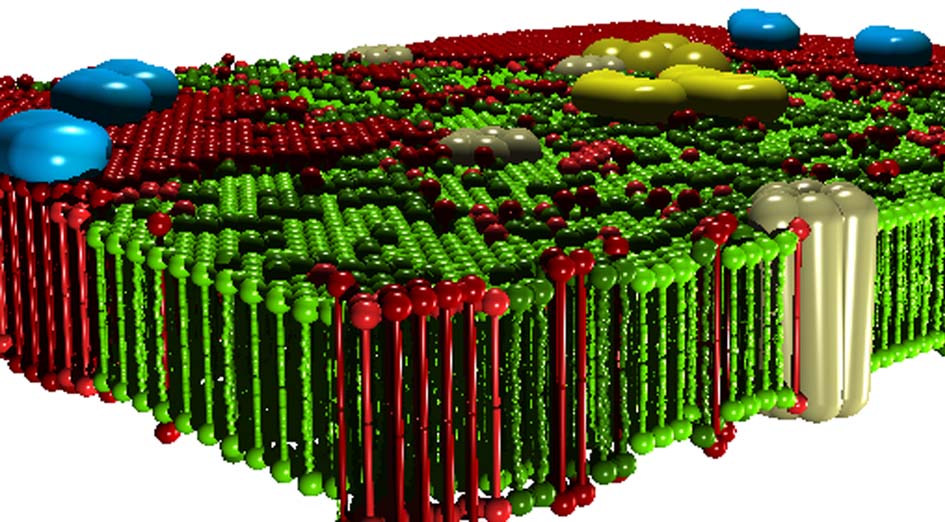}
 \parbox[c]{8cm}{ \caption*{Nerve pulse conduction along a membrane (shown here in a 3D representation with a heterogeneous distribution of lipids and proteins) can so far only be described in an incomplete manner. }}
\label{Figure0a}
    \end{center}
\end{figure}
In this article we outline a new model for the propagation of nerve signals through membranes (see Info box "Composition of a membrane") that relies exclusively on classical thermodynamics and hydrodynamics, containing all thermodynamic variables. This model allows unifying many observations made on nerves. It also explains the phenomenon of anesthesia whose mechanism has remained unexplained during the last 160 years. \\
\begin{figure}[hb!]
  \centering
  \fbox{\colorbox{hellorange}{
    \begin{minipage}{8.0 cm}
    \colorbox{navyblue}{\color{white}  \textbf{COMPACT}}
\begin{itemize}
  \item The presently accepted Hodgkin-Huxley model explains nerve pulse propagation by ion currents along concentration gradients.
  \item It only considers voltage dependent aspects and implies that heat is liberated during the pulse. However, this is not observed.
  \item Nerves pulses can not only be triggered by voltage but also by mechanical stimuli and local cooling.
  \item Many aspects of nerve pulse conduction can be derived from the cooperative properties of membranes. Close to their melting transition solitons can propagate. This picture describes more aspects of nerve conduction.
\end{itemize}
    \end{minipage}
  }
  }
\end{figure}

The propagation of the nervous impulse is one of the oldest problems in biophysics. In the late 18$^{th}$ century Luigi Galvani first described the contraction of a frog muscle after connecting two electrodes to the spine and the leg. He attributed the contraction to some kind of 'animal electricity' specific for life, a view distinctively different from the views of Schr\"odinger. In contrast, his contemporary Alessandro Volta was of the opinion that the pulse propagation is a purely electrical phenomenon. Although Galvani's view dominated the discussion at his time, we nowadays rather lean towards Volta's view. In the middle of the 19$^{th}$ century Hermann Helmholtz was the first to measure the velocity of a nerve pulse quantitatively. Interestingly, Helmholtz started his career as a physiologist and only later became the famous theoretical physicist. In his dissertation Helmholtz found that the pulse velocity in frog sciatic nerves is about 30 m/s. This occurred practically simultaneously with his formulation of the first law of thermodynamics. Today, it is practically unthinkable to find careers that span the whole range from physiology to theoretical physics, which is probably not to the advantage of Science in general.\\
\begin{figure}[htb]
  \centering
  \fbox{\colorbox{hellorange}{
    \begin{minipage}{8.0 cm}
    \colorbox{navyblue}{\color{white}  \textbf{COMPOSITION OF A BIOMEMBRANE}}\\*[0.2cm]
      \includegraphics[width=8cm]{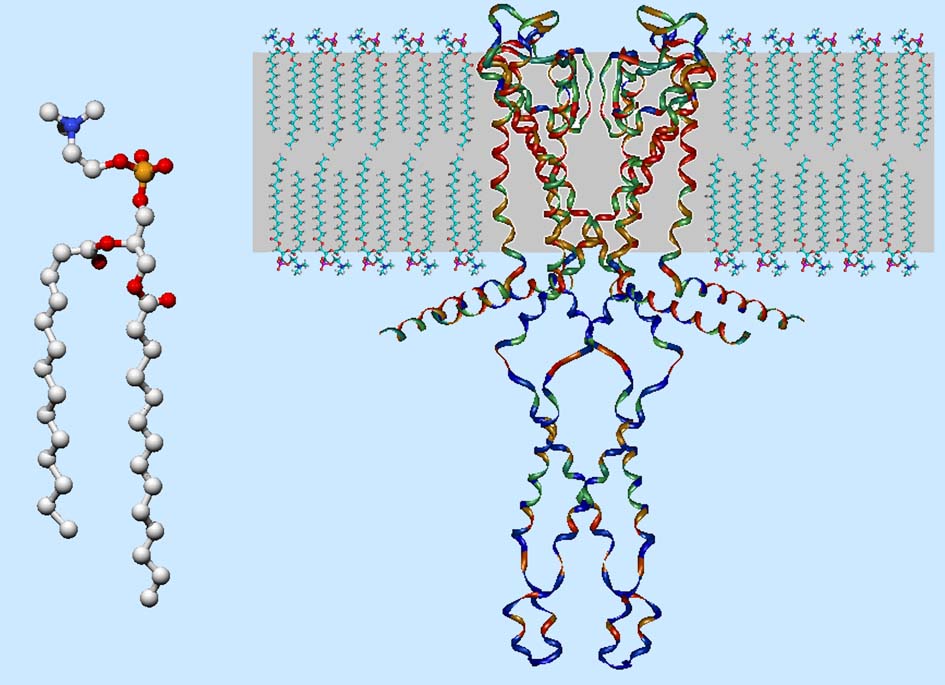}\\*[0.2cm]
      \parbox[c]{8cm}{ \caption*{{A biological membrane mainly consists of two classes of molecules. These are the lipids (left) and proteins that are embedded into the membrane (right) or associated to its surface (shown on the right is the potassium channel). Lipids are amphiphilic molecules that like water on one side and repel water on the other side due to the hydrocarbon chains. In contact to water lipids spontaneously form double layers (see Figure on page 1) that are the basic building blocks of biomembranes.}}}
      \label{Figure0}
    \end{minipage}
  }
  }
\end{figure}
\setcounter{figure}{0}

\textbf{\large Model with limitations }\\*[0.2cm]
The propagating electrical phenomena in nerves are called 'action potentials'. During a typical nerve pulse the voltage between the interior and the exterior of a cell changes locally by about 100mV. The contemporary view of these phenomena originates from the model of Alan L. Hodgkin and Andrew F. Huxley from 1952, for which they received the Nobel Prize in Medicine in 1963. In this model they attribute the voltage changes in nerve membranes to currents of sodium and potassium ions through individual proteins called ion channels (actually, in their original publication they considered conductances in a general manner). These channel proteins were characterized in the 1970s by Neher \& Sakmann with the patch clamp technique that they had developed. For this work they were awarded the Nobel Prize in Physiology in 1991. The role of the channel proteins during the conduction of the nervous impulse is thought to be the following: Upon changes in membrane potential, pores inside the proteins open in a complicated voltage- and time-dependent manner. The quantized opening and closing of such channels is shown in Fig. \ref{Figure6}, top. Potential changes caused by the ion fluxes influence the conductances of the channel proteins in their environment. This causes a cascade effect leading to the propagation of the voltage change. The physical description of the nerve pulse in this pictures rests on closed Kirchhoff current circuits. The proteins play the role of electrical resistors and the lipid membrane represents a capacitor. More details are given in the Infobox "The Hodgkin-Huxley model". This model only addresses voltage-related aspects of the nerve pulse. A fundamental characteristic of this view is that it rilies on dissipative processes and that it is therefore intrinsically of dissipative nature.\\  

However, the Hodgkin-Huxley picture has clear limitations. First of all, no explanation for the complex voltage- and time-dependence of the channel proteins is given. Instead, it is parametrized. This in itself demonstrates that the Hodgkin-Huxley model does not represent a theory in the strict physical sense. It does not have predictive power but is rather an \textit{a posteriori} description of the measurements. Furthermore, it restricts the measurement and the thinking one-sidedly on electrical phenomena. If the variation of the internal energy of the membrane is written as:
\begin{equation}
dE=T dS-p dV + ... + \Psi dq + ... + \sum_i \mu_i dn_i \,,
\end{equation}
the Hodgkin-Huxley picture only contains the $\Psi dq$ term describing the work necessary to charge the membrane capacitor. Not contained are changes in all other quantities, e.g., heat changes ($TdS$) or the work performed by variations in length and thickness. Careful measurements show, however, that all other thermodynamic quantities also vary in phase with the voltage changes across the membrane. This is not widely known - partially because such measurements require significantly more effort than recording voltage changes which is very simple. The accepted nerve model has no language for non-electrical phenomena. However, numerous changes of nerves are known that occur during the action potential.\\
In the following we focus on measurements of mechanical variations and reversible heat release in nerves.\\
\begin{figure*}[htb]
  \fbox{\colorbox{hellorange}{
    \begin{minipage}{17 cm}
    \colorbox{navyblue}{\color{white}  \textbf{THE HODGKIN-HUXLEY MODEL}}
    \begin{center}
    \includegraphics[width=14cm]{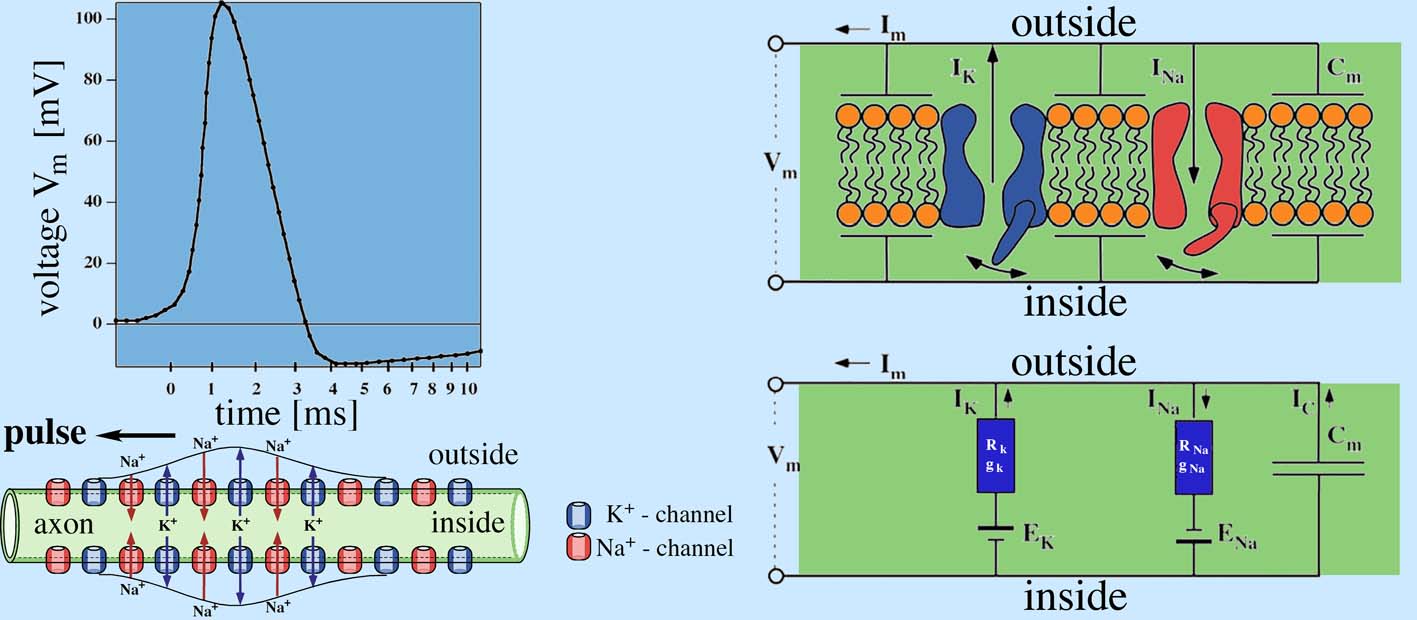} \\*[0.2cm]
      \parbox[c]{15cm}{ \caption*{{The basic elements of the Hodgkin-Huxley model. The nerve pulse from the original publication of Hodgkin and Huxley (top left) \cite{Hodgkin1952} is generated by the transient opening of sodium and potassium channels (bottom). Sodium and potassium ions flow through the channel proteins that are embedded in the membrane (top right). The corresponding electrical currents can be represented by an equivalent electrical circuit diagram (bottom right).  Here, $V_m$ is the observed membrane voltage. The quantities $I$ are the ion currents, the $E_{Na}$ and $E_K$ are the Nernst potentials of sodium and potassium. $C_m$ is the membrane capacitance. Adapted from \cite{Heimburg2007b}.}}}\\*[0.1cm]
      \label{Figure1}
      \end{center}
The Hodgkin-Huxley model is based on the following assumptions: Within the nerve cell the potassium concentration is high, while outside of the cell one finds a high sodium concentration. The concentration differences cause a voltage difference between inside and outside that is related to the Nernst potentials. Embedded into the membrane one finds channel proteins that are specifically conducting either sodium or potassium ions. After a perturbation caused by voltage changes, these proteins become conductive for the respective ions and ion currents flow through the membrane. The details of this process are not understood. It has been assumed that charged groups within the protein move in the external field. The membrane is considered as a capacitor that becomes transiently charged. Local voltage changes open channel proteins in the environment. This process leads to a cascade effect. The nerve signal itself consists of the moving segment of charged membrane capacitor. The Hodgkin-Huxley model is of dissipative nature because it relies on ion currents along concentration gradients.
    \end{minipage}
  }
  }
\end{figure*}

\begin{figure}[t!]
    \begin{center}
	\includegraphics[width=8.5cm]{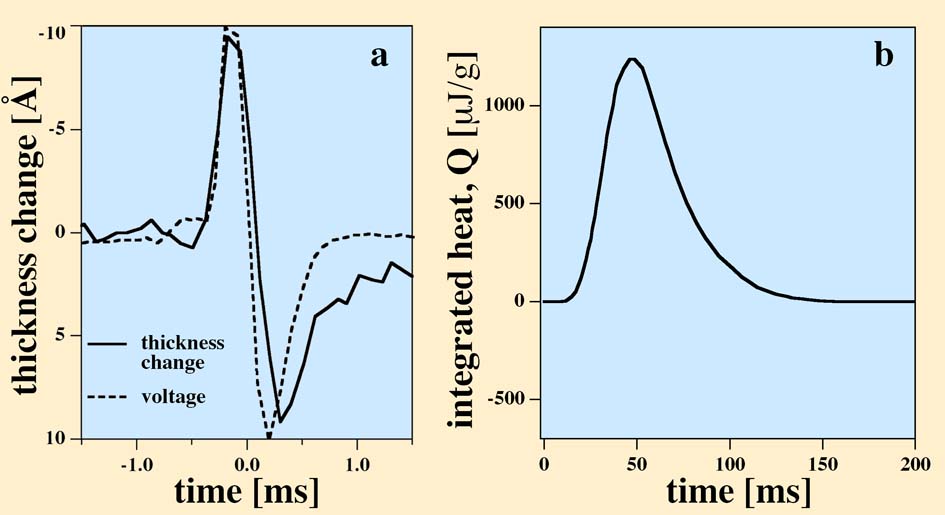}\\*[0.2cm]
	\parbox[c]{8cm}{ \caption{{During the nerve pulse both thickness and heat change. In the quid axon heat and voltage change in phase (a). The voltage scale was adjusted to show the proportionality  \cite{Iwasa1980b}. The heat release rate of an olfactory pike nerve first displays a positive and then a negative value (b). Integrated over the pulse the overall heat release is approximately zero \cite{Ritchie1985}. }
	\label{Figure2_3}}}
    \end{center}
\end{figure}

\textbf{\large Nerve excitation}\\*[0.2cm]
Nerve signals can not only be triggered by voltage changes. The observed excitability by mechanical stimuli caused E. Wilke already in 1912 \cite{Wilke1912b} to conclude that  the nervous impulse cannot just be a purely electrical event. Instead, he proposed that it is a piezoelectric phenomenon. He documented the mechanical changes in a simple experiment: Presuming that nerves  simultaneously become thicker and shorter during excitation, Wilke connected a thin glass fiber to the end of a nerve and showed that it started vibrating during nerve excitation. Ever since then the possibility of mechanical changes has been discussed by various renowned authors, among them Curtis \& Cole, Hodgkin \& Huxley and later in particular by Tasaki. The latter author demonstrated in a series of studies that nerves in fact not only do contract but that they also change their thickness (by about 1nm, see Fig.  \ref{Figure2_3}a). One can measure forces with a cantilever (order of magnitude 1 nanonewton) that the nerves exert on the environment during nerve excitation.\\*[0.2cm]
The excitability of nerves by local cooling is another remarkable observation. It indicates that nerve excitation is related to heat changes. This was first investigated by Helmholtz in 1847 - however,  with yet insufficient methods. The first reliable measurements of heat production in nerves stem from Archibald V. Hill. For his research on the heat production in muscles he received the Nobel Prize of Medicine in 1922. He extensively studied heat exchange of nerves as well.  His results are summarized in a review from 1958 \cite{Abbott1958}: During the first phase of the nervous impulse heat is liberated while it is mostly reabsorbed in a second phase. If the heat production is integrated over the duration of the nerve pulse one obtains a total heat of zero within the resolution of the experiment (Fig. \ref{Figure2_3}b). This implies that the nervous impulse has to be considered as an adiabatic and reversible process. The heat liberation is in phase with the voltage changes. This result that was reproduced by a number of other authors is in conflict with the Hodgkin-Huxley model, which as indicated above is of dissipative nature. In the Hodgkin-Huxley model ion currents flow through proteins that are modeled as resistors. Taking this model literally one should find heat liberation during the nervous impulse. The different direction of the sodium and potassium currents should not play a role. However, this is obviously not what is found in experiments on nerves.\\ 

\textbf{\large Size matters}\\*[0.2cm]
One of the big triumphs of modern biology is the discovery of the DNA double helix by Watson and Crick and the subsequent unraveling of the genetic code (following the proposals of the nuclear physicist George Gamov). As is well-known the DNA contains segments called genes that encode the structure of proteins. These proteins possess various functions in transport and especially in catalysis. It is generally assumed that the major part of biological activity of cells depends on the function of these proteins, which are either functional as single molecules or as rigid complexes of such molecules. Often, such proteins are called nano-machines. The presently popular focus on single molecules is probably often a reasonable assumption. However, it has the disadvantage that the view on the thermodynamics of such systems is lost. Even though the principles of thermodynamics are valid for the isolated molecule, the real strength of thermodynamics lies in the collective phenomena. Temperature, pressure, heat, and so forth, are terms from the world of ensembles. The second law of thermodynamics is valid for the complete system under investigation and not for arbitrarily chosen parts of this system, and hence not necessarily for the single molecule within a cell. Thus, the success of molecular genetics has in some respect  contributed to the fact that classical physics is practically not considered any longer in biology. It seems instead as if the opinion has prevailed that one can understand life if one understands all single molecules. We do not share this opinion.\\

A common approach in physics is to search for explanations on length scales that are characteristic of the phenomenon one wishes to observe. For instance, it is not meaningful trying to understand the propagation of sound on the atomic level. Typically, sound is described by a very simple differential  equation that contains the sound velocity (or the compressibility on the scale of the wave length) as the only parameters. While this parameter displays slightly different values in helium than in air, the physics of the propagation phenomenon is independent of this. One could also say that the molecular detail does not play a fundamental role for the physics of the phenomenon. Quite many processes in physics are of a nature that does not require knowledge of atomic or molecular processes. This is especially true for cooperative transitions, e.g., phase transitions like the melting of ice or the melting of membranes (see next section).\\
A typical signal in a nerve lasts about a millisecond and propagates with about 100 m/s in motor neurons. Against intuition this means that a nerve pulse is several centimeters long. It is in no respect a microscopic phenomenon. In contrast, the diameter of a channel protein is only about 5 nm, which is more than seven orders of magnitude smaller. The size difference is comparable to that between a saucer and the diameter of the continent of Europe. However, one would typically not exclusively ascribe processes that happen on the length scale of Europe (e.g., weather and tectonic plate movement) to objects of the size of a saucer. But this is exactly what happens in the models for nerve pulse propagation (and generally for many models in biology). They are nearly exclusively of molecular nature.\\

\begin{figure}[b!]
    \begin{center}
	\includegraphics[width=8.5cm]{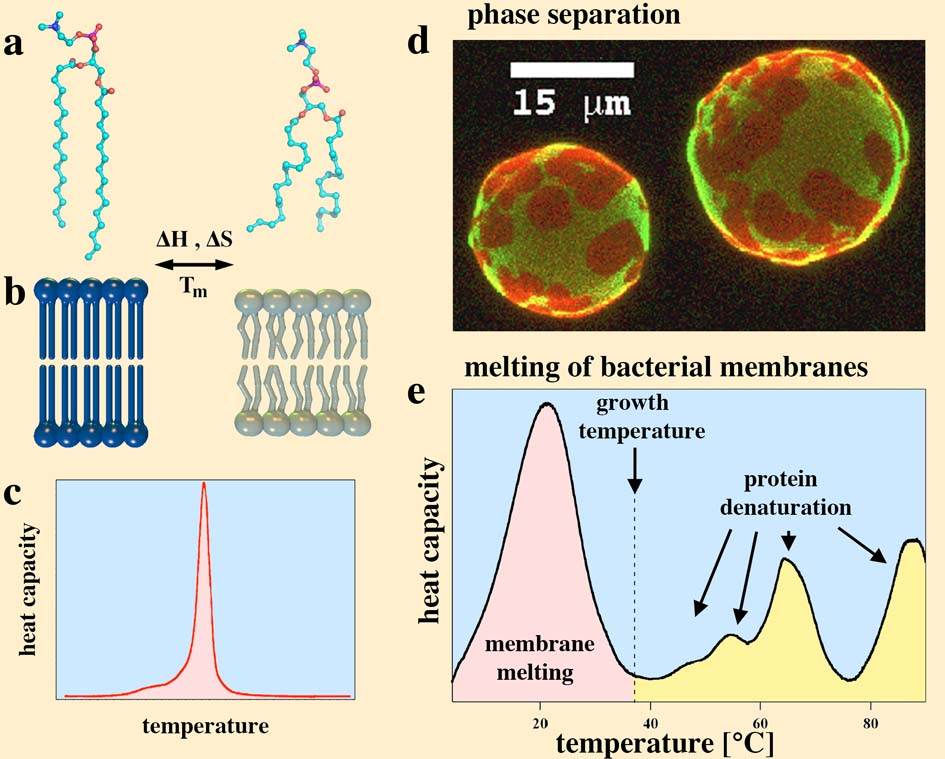}\\*[0.2cm]
	\parbox[c]{8cm}{ \caption{{
	During melting the  internal order of lipids changes (a) and the membranes becomes liquid (b). However, it stays intact during melting. At the transition temperature the susceptibilities, e.g. the heat capacity, are at maximum (c, shown for a synthetic lipid membrane). Phase separation in the melting regime of vesicles made from a lipid mixture can be visualized with fluorescence microcopy (d). The red and green regions show solid and liquid domains. Below the growth temperature of bacterial membranes at 37$^{\circ}$C the heat capacity displays a broad melting regime of the membrane (e). Above  growth temperature several proteins denature. }
	\label{Figure4}}}
    \end{center}
\end{figure}
\textbf{\large Membrane melting}\\*[0.2cm]
In the following we will derive many aspects of nerve pulse propagation from the cooperative properties of membranes. In many biological membranes one finds order transitions of the membrane lipids about 10 degrees below body temperature. In simple terms one could say that at a temperature $T_m$ the membranes melt. During melting the lateral arrangement of the lipid molecules changes, and so does the intramolecular order of the lipid chains (Fig. \ref{Figure4}a-c). Both the enthalpy ($\Delta H$) and the entropy ($\Delta S$) increase and heat is absorbed. One has to envision this as a melting in two dimensions that leaves the structure of the membrane intact - it still separates different aqueous compartments. In contrast to chemically pure lipids  (Fig. \ref{Figure4}c) transitions in biomembranes are relatively broad (Fig. \ref{Figure4}e) because they consist of a complex mixture of hundreds of different lipid species with different properties. As a consequence of the fluctuation-dissipation theorem many physical properties change within these transitions. Examples are the heat capacity, the compressibilities, the bending stiffness and the relaxation times. Furthermore, membranes become very permeable for ions and molecules. Within the melting regime one also finds phase separation and domain formation indicating large fluctuations in state and composition. This can nicely be visualized by fluorescence microscopy (Fig. \ref{Figure4}d).\\
The position of the transition responds sensitively to changes of the intensive variables such as pressure, temperature and pH (the chemical potential of protons). Hence, there is the possibility of altering the susceptibilities of membranes by changing the intensive variables \cite{Heimburg1998, Heimburg2007c}.
\begin{figure*}[htb!]
    \begin{center}
	\includegraphics[width=15cm]{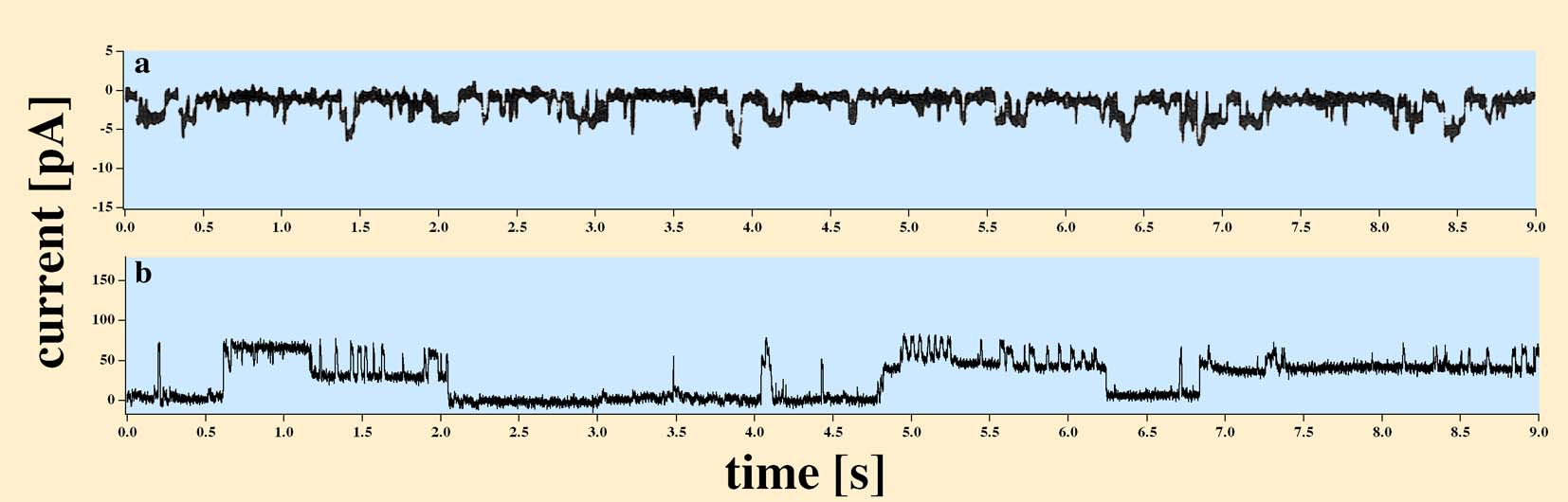}\\*[0.2cm]
	\parbox[c]{15cm}{ \caption{{(a) Currents through membranes of a frog muscle cell containing the acetylcholine receptor are quantized \cite{Neher1976}. (b) This is also true for currents through a synthetic lipid membrane close to the transition maximum \cite{Blicher2009}. In spite of slightly different experimental conditions, amplitude and typical time scale are similar. Multiple steps in the data indicate that several pores might open simultaneously. }
	\label{Figure6}}}
    \end{center}
\end{figure*}
For instance, by this effect one can change the permeability of membranes by shifting the melting events through adjustments of temperature, pressure or pH. The generation of a membrane pore requires performing work. The softer the membranes the easier it is to form pores by thermal fluctuations. Empirically one finds that the membrane conductance is proportional to the heat capacity. As a consequence, the conductance depends in a predictable manner on changes of the thermodynamic variables.\\
When measuring currents through small membrane segments, the conductance events appear in a quantized steps, i.e. one observes a switching between current-on and current-off events. In biological membranes one normally assumes that this phenomenon is caused by the opening and closing of ion channel proteins (Fig. \ref{Figure6}a). The two conduction states are called open and closed channels. As already described above, these channel proteins play an important role in the Hodgkin-Huxley model of nerve pulse conduction. Interestingly, one finds very similar events in synthetic membranes that are free of proteins (e.g., \cite{Antonov1980, Blicher2009}), in particular if one is close to the melting transition (Fig. \ref{Figure6}b). Both current amplitude and the typical opening time scales are practically indistinguishable from those of protein-containing membranes. Obviously, the finding of quantized current events is not a proof for the activity of proteins even though this is what is generally assumed since membranes close to transition display the same characteristics. Whether the events in protein-containing membranes and those in pure lipid membranes have the same origin is yet an open question.\\

\textbf{\large A pulse along the membrane}\\*[0.2cm]
The above considerations about reversible changes of heat and thickness during the nerve pulse and the quantized conductance events of the membranes close to transitions have let us to consider the nerve pulse as a reversible and cooperative phenomenon. Under physiological conditions biological membranes are close to order transitions that are associated to large changes in the elastic constants. We have recently shown that under these conditions one expects the propagation of electromechanical solitons along membrane cylinders (as are the axons of nerves) \cite{Heimburg2005c, Heimburg2007b}. During this pulse the membrane is transiently moved through the transition while reversibly releasing the heat of melting. For this reason one can trigger a pulse by local cooling. \\
Solitons are localized excitations that propagate without attenuation or change in shape. They were first found by the ship engineer John Scott Russell. Riding a horse, he followed a localized water wave on the Union Canal close to Edinburgh. In the following, he described this phenomenon both experimentally and theoretically. The conditions for the existence of solitons are the presence of both nonlinearity of the speed of sound as a function of amplitude, and dispersion which is the frequency dependence of the sound velocity. Exactly these properties are found in membranes close to transitions. Such pulses rely on reversible mechanics meaning that they are not dissipative phenomena. One expects a reversible heat release consisting of the heat of melting. Since membranes carry electrical charges, such a soliton has electrical properties. Therefore, one can consider the pulses as piezoelectric pulses.\\
\begin{figure}[tb!]
    \begin{center}
	\includegraphics[width=8cm]{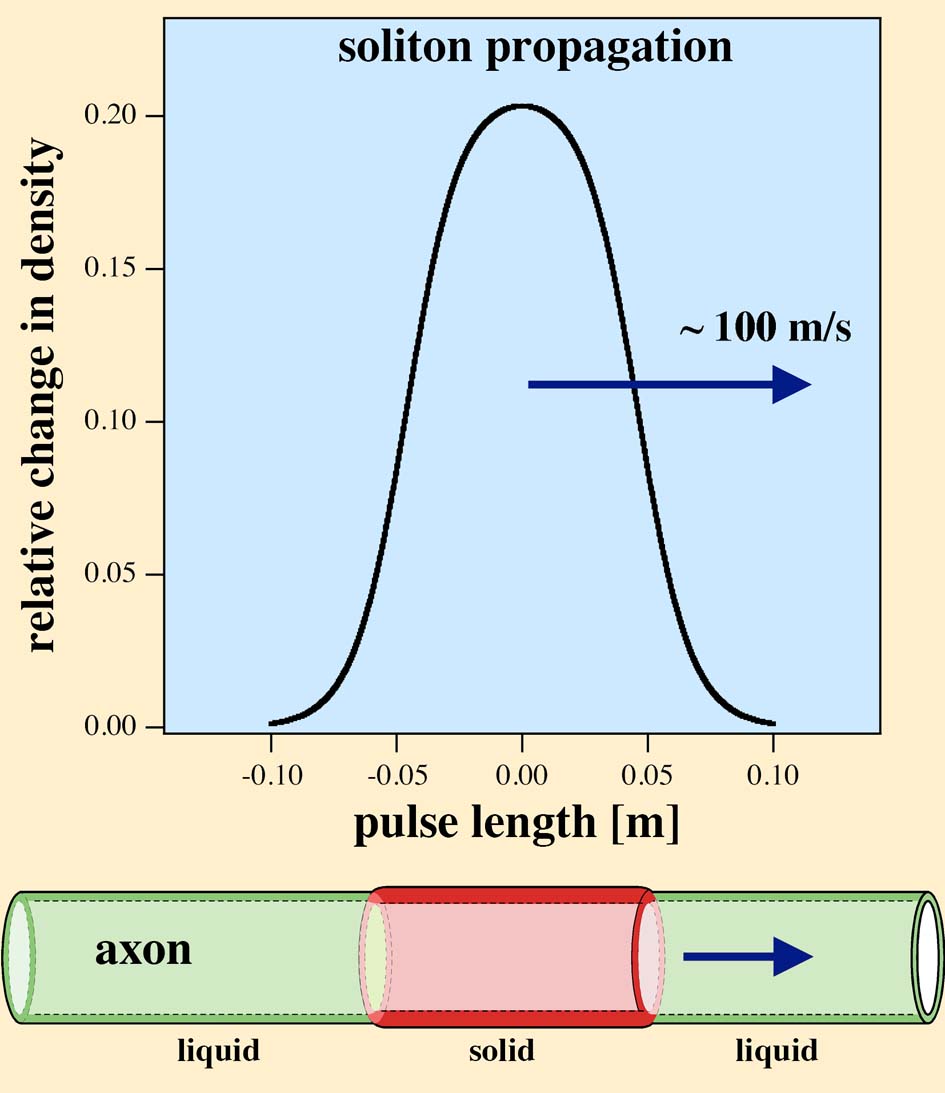}\\*[0.2cm]
	\parbox[c]{8cm}{ \caption{{In the proximity of the melting transition the elastic constants are nonlinear functions of the lateral density. This allows for the formation of stable localized  pulses. The lateral density in these pulses is increased by about 20\%. This corresponds to the density change from liquid to solid. The solitons propagate with a velocity close to the lateral speed of sound of lipid membranes, ie., with about 100 m/s. This corresponds to the pulse velocity of myelinated nerves. }
	\label{Figure5}}}
    \end{center}
\end{figure}
For the description of nerves this point of view has a number of advantages:
\begin{itemize}
   \item It explains the reversible heat release, and the thickness and length changes. 
 \item It explains the excitability of nerves by so different variables as voltage, temperature and mechanical stimulation. Each thermodynamic variable corresponds to an aspect of the signal.
 \item It has predictive power because the influence of many variables on the cooperative events in membranes is known and follows the laws of thermodynamics.
 \item During the pulse propagation one finds quantized currents through the membranes.
 \item As explained in the following, it also explains many aspects of anesthesia.
\end{itemize}

\textbf{\large The action of anesthetics}\\*[0.2cm]
To excite the electromechanical pulses described in the previous section one has to move the membrane locally through its melting transition (e.g., by locally lowering the temperature). The necessary free energy is proportional to the distance of the transition from the physiological temperature. Every change of a thermodynamic variable that leads to a change in melting temperature also influences the excitability of the membrane. Among those variables are hydrostatic pressure, pH (concentration of protons), calcium concentration, and the change of the concentration of membrane soluble substances. A very interesting effect is found for the anesthetics. \\
The action of anesthetizing substances has been know for a very long time. In 1846, the physician William Morton from the Massachusetts General Hospital in Boston (Harvard) publicly displayed the action of diethyl ether during an operation. Ever since then one has found many substances with very different chemical structures that cause anesthesia, among them laughing gas, halothane, chloroform and many alcohols, but also the noble gas xenon. All of these anesthetics follow the famous Meyer-Overton rule that states that the anesthetic potency is exactly proportional to their solubility in membranes \cite{Overton1901}. This law is valid over several orders of magnitude of the solubility from laughing gas to long chain alcohols. This implies that at the critical anesthetic dose ED$_{50}$ where 50\% of the individuals loose conscience one always finds exactly the same concentration of anesthetics in membranes, completely independent of their chemical nature. This correlation has never really been explained. In his famous book from 1901, Overton himself proposed searching for general and unspecific physicochemical explanations.   \\
Interestingly, it has also been shown that anesthetics influence the melting points of membranes  \cite{Heimburg2007c}. The melting point depression $\Delta T_m$ is well described by the well-known law  $\Delta T_m=(RT_m^2/\Delta H)\cdot x_A$, where  $\Delta H$ is the heat of melting and $x_A$ is the molar fraction of anesthetics in the membrane. The very same law also describes the effect of salt on the melting of ice. As found experimentally for the anesthetics, this law is independent of the chemical nature of the anesthetics. It only requires the ideal solubility of the molecules in the liquid membranes. One can therefore rephrase the Meyer-Overton correlation in the following manner: The critical anesthetic dose ED$_{50}$ is exactly proportional to their ability to lower phase transitions. As a consequence the membranes are less excitable. Furthermore, the anesthetics alter the membrane permeability. We have recently shown that one can 'block' ion channel events in pure lipid membranes with anesthetics \cite{Blicher2009}.
While anesthetics lower melting points, hydrostatic pressure cause an increase of the melting point due to the latent volume of the transition. If there were any relation between melting points and anesthesia, hydrostatic pressure should be able to reverse the anesthetic effect. This was in fact found experimentally (e.g., \cite{Johnson1950}). Anesthetized tadpoles returned to normal activity at a pressure of about 50 bars. This is exactly the order of magnitude that one would predict from the pressure dependence of the membrane transitions. A similar correlation is found for pH decreases \cite{Heimburg2007c}.\\

\textbf{\large Summary}\\*[0.2cm]
We wanted to demonstrate that simple physical methods can help understanding phenomena in biological membranes without knowledge of molecular detail. This implies the excitability of membranes, their permeability, anesthesia and its reversal by hydrostatic pressure. Summarizing, this means that membrane excitability depends on all thermodynamic variables in a predictable manner that is described by a single potential: The free energy of the membrane. Due to their chemical potentials, the membrane proteins play a thermodynamical role and are themselves thermodynamic variables.\\ 

\textbf{Acknowledgment:} I am grateful to Dr. K. Kaufmann from G\"ottingen for introducing me to this interesting subject. He himself has worked on many of these phenomena. Thanks to A. Blicher and K. Feld for proofreading my translation.

\footnotesize{

}
\begin{figure}[htb]
  \fbox{\colorbox{hellorange}{
    \begin{minipage}{8 cm}
    \colorbox{navyblue}{\color{white}  \textbf{THE AUTHOR}}
\begin{center}
\hfill\includegraphics[width=3.8cm]{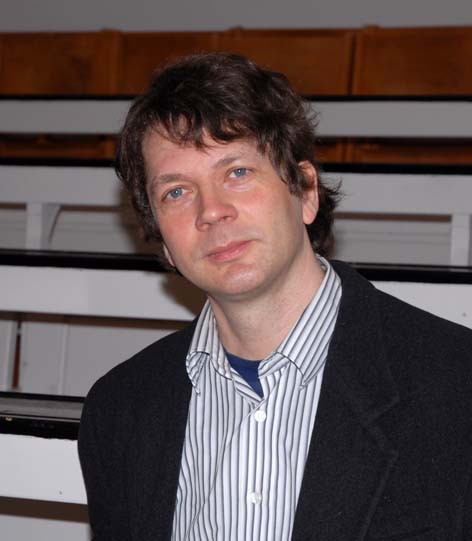}\\*[0.2cm]
\end{center}  
Thomas Heimburg (*1960)  studied Physics in Stuttgart and G\"ottingen, Germany. He obtained his PhD in 1989 at the Max Planck Institute for Biophysical Chemistry in     G\"ottingen. After being a postdoctoral fellow at the University of Virginia he got his Habilitation degree in the field of Biophysics at the University of G\"ottingen in 1995. From 1997-2003 he lead an independent reseach group at the MPI for Biophysical Chemistry. Since 2003 he is Associate Professor for Biophysics at the Niels Bohr Institute of the University of Copenhagen, where he has been the head of the \textit{Membrane Biophysics Group}. 
\label{Figure8}
    \end{minipage}
  }
  }
\end{figure}


\begin{thebibliography}{10}

\bibitem{Schroedinger1944}
Schr\"odinger, E., 1944.
\newblock What is life?
\newblock Cambridge University Press.

\bibitem{Hodgkin1952}
Hodgkin, A.~L., and A.~F. Huxley.
\newblock 1952.
\newblock A quantitative description of membrane current and its application to
  conduction and excitation in nerve.
\newblock J.\ Physiol. 117:500--544.

\bibitem{Heimburg2007b}
Heimburg, T., and A.~D. Jackson.
\newblock 2007.
\newblock On the action potential as a propagating density pulse and the role
  of anesthetics.
\newblock Biophys. Rev. Lett. 2:57--78.

\bibitem{Wilke1912b}
Wilke, E., and E.~Atzler.
\newblock 1912.
\newblock Experimentelle {B}eitr{\"a}ge zum {P}roblem der {R}eizleitung im
  {N}erven.
\newblock Pfl\"ugers Arch. 146:430--446.

\bibitem{Ritchie1985}
Ritchie, J.~M., and R.~D. Keynes.
\newblock 1985.
\newblock The production and absorption of heat associated with electrical
  activity in nerve and electric organ.
\newblock Quart.\ Rev.\ Biophys. 392:451--476.

\bibitem{Iwasa1980b}
Iwasa, K., I.~Tasaki, and R.~C. Gibbons.
\newblock 1980.
\newblock Swelling of nerve fibres associated with action potentials.
\newblock Science 210.

\bibitem{Abbott1958}
Abbott, B.~C., A.~V. Hill, and J.~V. Howarth.
\newblock 1958.
\newblock The positive and negative heat production associated with a nerve
  impulse.
\newblock Proc. R. Soc. London. B 148:149--187.

\bibitem{Heimburg1998}
Heimburg, T.
\newblock 1998.
\newblock Mechanical aspects of membrane thermodynamics. {E}stimation of the
  mechanical properties of lipid membranes close to the chain melting
  transition from calorimetry.
\newblock Biochim.\ Biophys.\ Acta 1415:147--162.

\bibitem{Heimburg2007c}
Heimburg, T., and A.~D. Jackson.
\newblock 2007.
\newblock The thermodynamics of general anesthesia.
\newblock Biophys.\ J. 92:3159--3165.

\bibitem{Antonov1980}
Antonov, V.~F., V.~V. Petrov, A.~A. Molnar, D.~A. Predvoditelev, and A.~S.
  Ivanov.
\newblock 1980.
\newblock The appearance of single-ion channels in unmodified lipid bilayer
  membranes at the phase transition temperature.
\newblock Nature 283:585--586.

\bibitem{Blicher2009}
Blicher, A., K.~Wodzinska, M.~Fidorra, M.~Winterhalter, and T.~Heimburg.
\newblock 2009.
\newblock The temperature dependence of lipid membrane permeability, its
  quantized nature, and the influence of anesthetics.
\newblock Biophys.\ J. 96: 4581-4591.

\bibitem{Neher1976}
Neher, E., and B.~Sakmann.
\newblock 1976.
\newblock Single-channel currents recorded from membrane of denervated frog
  muscle fibres.
\newblock Nature 260:779--802.

\bibitem{Heimburg2005c}
Heimburg, T., and A.~D. Jackson.
\newblock 2005.
\newblock On soliton propagation in biomembranes and nerves.
\newblock Proc.\ Natl.\ Acad.\ Sci.\ USA 102:9790--9795.

\bibitem{Overton1901}
Overton, C.~E., 1901. Jena, Germany. 
\newblock Studien \"{u}ber die {N}arkose.
\newblock Verlag Gustav Fischer.
\newblock {E}nglish translation: {S}tudies of
  {N}arcosis, {C}hapman and {H}all, 1991, {R.} {Lipnick, Ed.}

\bibitem{Johnson1950}
Johnson, F.~H., and E.~A. Flagler.
\newblock 1950.
\newblock Hydrostatic pressure reversal of narcosis in tadpoles.
\newblock Science 112:91--92.

\end{thebibliography}
\end{document}